\documentclass{svjour3}                     % onecolumn (standard format)
\smartqed  % flush right qed marks, e.g. at end of proof
\usepackage{color,graphicx,hyperref,amsmath,amssymb,mathrsfs}
\usepackage{ulem}%to cross text

\renewcommand{\Re}{\operatorname{Re}}
\renewcommand{\Im}{\operatorname{Im}}

\begin{document}

\title{Bifurcations in the time-delayed Kuramoto model of coupled
oscillators: Exact results}
\titlerunning{Bifurcations in the time-delayed Kuramoto model of coupled
oscillators}        % if too long for running head

\author{David M\'{e}tivier        \and
        Shamik Gupta}

\institute{David M\'etivier \at
              Center for Nonlinear Studies and Theoretical Division T-4 of Los Alamos National Laboratory, NM 87544, USA \\
              Tel.: +1 (505) 667-7052\\
              \email{metivier@lanl.gov}           %  \\
           \and
           Shamik Gupta \at
              Department of Physics, Ramakrishna Mission Vivekananda
              University, Belur Math, Howrah 711202, India \\
\email{shamik.gupta@rkmvu.ac.in}
}

\date{Received: date / Accepted: date}

\maketitle

\begin{abstract}
In the context of the Kuramoto model of coupled oscillators with distributed natural frequencies interacting through a time-delayed mean-field, we derive as a function of the delay exact results for  the stability boundary between the incoherent and the synchronized state and the nature in which the  latter bifurcates from the former at the critical point. Our results are based on an unstable manifold expansion in the vicinity of the bifurcation, which we apply to both the kinetic equation for the single-oscillator distribution function in the case of a generic frequency distribution and the         corresponding Ott-Antonsen(OA)-reduced dynamics in the special case of a Lorentzian distribution. Besides elucidating the effects of delay on the nature of bifurcation, we show that the approach due to Ott and Antonsen, although an ansatz, gives an amplitude dynamics of the unstable modes close to the bifurcation that remarkably coincides with the one derived from the kinetic equation. Further more, quite interestingly and remarkably, we show that close to the bifurcation, the unstable manifold derived from the kinetic equation has the same form as the OA manifold, implying thereby that the OA-ansatz form follows also as a result of the unstable manifold expansion. We illustrate our results by showing how delay can affect dramatically the bifurcation of a bimodal distribution. 
\end{abstract}

%%%%%%%%%%%%%%%%%%%%%%%%%%%%%%%%%%%%%%%%%%%%%%%%%%%%%%%%%%%%%%%%%%%%%%%%%%%%%%%%%%%%%
\section{Introduction}

%\dm{
%
%
%Starting from the aforementioned developments and with an aim
%to answer the questions just raised, we embark here on a detailed
%analytical characterization of bifurcation in~\eqref{eq:eom-1}. It has
%been two decades since the introduction of model~\eqref{eq:eom},
%and despite its widespread discussions (some recent
%ones are Refs.~\cite{YS-app6,YS-app5,YS-app4,YS-app3,YS-app2,YS-app1})), it is
%here that we report its complete analytical characterization, thereby
%solving an outstanding problem in nonlinear sciences. 
%
%
%
%The success of the OA approach in reducing dynamical
%complexity led to hundreds of applications in
%applied mathematics and physics; some recent ones are
%Refs.~\cite{Ott-app8,Ott-app7,Ott-app6,Ott-app5,Ott-app4,Ott-app3,Ott-app2,Ott-app1}. 
%}

The Kuramoto model enjoys a unique status in the field of nonlinear
dynamics~\cite{Kuramoto:1984,Strogatz:2000,Acebron:2005,Gupta:2014-2,Rodrigues:2016,Gherardini:2018,Gupta:2018}.
It provides arguably the minimal framework to model the phenomenon of spontaneous
synchronization commonly observed in
nature~\cite{Pikovsky:2001,Strogatz:2003}, for example, among groups of fireflies flashing on and off in
unison~\cite{Buck:1988}, in cardiac pacemaker cells~\cite{Peskin:1975},
in electrochemical~\cite{Kiss:2002} and electronic~\cite{Temirbayev:2012} oscillators, in Josephson
junction arrays~\cite{Benz:1991} and in electrical power-grid
networks~\cite{Rohden:2012}, to name a few. Modeling the individual
units in a synchronizing system as nearly-identical limit-cycle oscillators, the Kuramoto
model arises on considering the individual limit-cycles to be interacting weakly with one another, with the strength of coupling being
the same for every pair of oscillators~\cite{Pikovsky:2001,Gupta:2018}. 
Specifically, the model comprises a set of $N$ oscillators with
distributed natural frequencies $\omega_j \in [-\infty,\infty];~j=1,2,\ldots,N$. Denoting
by $\theta_j \in [0,2\pi)$ the phase of the $j$-th oscillator, the dynamics of the model is given by a set of $N$
coupled equations of the form \cite{Kuramoto:1984}
\begin{equation}
\dfrac{{\rm d}\theta_j(t)}{{\rm d}t}=\omega_j + \dfrac{K}{N}\sum_{k=1}^N \sin
(\theta_k(t)-\theta_j(t)),
\label{eq:eom-kuramoto}
\end{equation}
where $K\ge 0$ is the coupling constant. The scaling of $K$ by $N$
ensures that the associated term is well behaved in the thermodynamic limit $N \to
\infty$, which is indeed the limit of interest to us in this paper and
which also models the fact that in this limit, the coupling between any pair of
oscillators is weak, scaling as $1/N$. The frequencies
$\{\omega_j\}_{1 \le j \le N}$ denote a set of quenched disordered random
variables distributed according to a common distribution ${\cal G}(\omega)$,
with the latter obeying the normalization $\int_{-\infty}^\infty {\rm
d}\omega~{\cal G}(\omega)=1$. 

The coherence among the phases of the oscillators in the Kuramoto model is
conveniently measured by the complex order parameter $r(t)$,
defined as
\begin{equation}
r(t)\equiv\dfrac{1}{N}\sum_{j=1}^N e^{i\theta_j(t)},
\label{eq:rt-definition}
\end{equation}
where the quantity $|r|$ measures the amount of
coherence. A perfectly synchronized state corresponds to $|r|=1$, while in an
incoherent state, when the phases $\theta_j$ are uniformly distributed
in $[0,2\pi)$, one has $|r|=0$. In terms of the order parameter, one may
rewrite the equation of motion~(\ref{eq:eom-kuramoto}) as
\begin{equation}
\dfrac{{\rm d}\theta_j(t)}{{\rm d}t}=\omega_j+K\Im \left[r(t)e^{-i\left(\theta_j(t)\right)}\right],
\label{eq:Kuramoto-eom-1}
\end{equation}
which makes it evident that the phase of
an oscillator at a given time evolves due to the effect
of the instantaneous mean field $r$ built up in the system from the
interaction of the oscillator with all the other oscillators.

The setting of the Kuramoto model is really the simplest one can
conceive. It has
indeed been very successful in addressing theoretically the emergence of synchrony in
diverse dynamical settings involving coupled oscillators, showing in
particular in the limit $N \to \infty$ and for a given distribution of natural frequencies
of the oscillators that the system exhibits a
spontaneous transition from incoherence to synchrony as
the coupling strength $K$ is increased beyond a
certain critical value~\cite{Kuramoto:1984,Strogatz:2000}. Nevertheless, one does
encounter situations where due to time delay in propagation of signals
between the interacting units, the evolution of the dynamical variables is
determined not entirely by their instantaneous values but has an
essential dependence also on the values of the dynamical variables at an earlier time instant. Time delay is known to have important
consequences, for example, for synchronization of biological
clocks~\cite{Oates:2010}, in networks of digital
phase-locked loops~\cite{Wetzel:2017}, and in the case of information propagation
through a neural network in which delay is known to influence the temporal
characteristics of oscillatory behavior of neural
circuits~\cite{Blondeau:1992}.  It has been
argued that presence of even a small time delay may affect in general the global
dynamics of ensembles of limit-cycle oscillators~\cite{Niebur:1991}.

To assess the effects of time delay, a pioneering work within the ambit of the
Kuramoto model was pursued by Yeung and Strogatz in
Ref.~\cite{Yeung:1999}, in which they
considered a generalization that allows for a 
time-delayed mean-field interaction between the oscillating units.
Specifically, in terms of a time delay $\tau >0$, the phases evolve in
time according to the dynamics
\begin{equation}
\dfrac{{\rm d}\theta_j(t)}{{\rm d}t}=\omega_j+\dfrac{K}{N}\sum_{k=1}^N
\sin\left[\theta_k(t-\tau)-\theta_j(t)-\alpha\right].
\label{eq:eom}
\end{equation}
In terms of $r(t)$,
the dynamics~(\ref{eq:eom}) takes the form
\begin{equation}
\dfrac{{\rm d}\theta_j(t)}{{\rm d}t}=\omega_j+K\Im \left[r(t-\tau)e^{-i\left(\theta_j(t)+\alpha\right)}\right],
\label{eq:eom-1-rt}
\end{equation}
which puts in evidence the presence of a delayed mean-field: the
time evolution of oscillator phases at time $t$ is affected by the
value of the mean-field $r(t-\tau)$ measuring the macroscopic state of
the system at a previous instant $t-\tau$. Here, $\alpha \in
(-\pi/2,\pi/2)$ is the so-called phase frustration
parameter~\cite{Sakaguchi:1986}.
Note that setting the phase
frustration parameter and the time 
delay to zero, $\alpha=0,\tau=0$, reduces the
dynamics~(\ref{eq:eom}) to that of the Kuramoto
model~(\ref{eq:eom-kuramoto}).

We may anticipate that introducing delay in the
Kuramoto model may lead to a richer and more
complex dynamical scenario, which would at the same time be theoretically more
challenging to analyze. Indeed, Ref.~\cite{Yeung:1999} unraveled a range of new phenomena including bistability between synchronized and incoherent states and unsteady solutions
with time-dependent order parameters that do not occur in the original
Kuramoto model~(\ref{eq:eom-kuramoto}). In the particular case of
oscillators with identical natural frequencies and with $\alpha=0$,
Yeung and Strogatz were able to derive
exact formulas for the stability boundaries between the
incoherent and synchronized states. For the general case of oscillators
with distributed natural frequencies, still with $\alpha=0$, they adduced numerical results for the case of a Lorentzian
frequency distribution to
suggest the occurrence of
bifurcation of the incoherent state as a function of $K$, which could be either sub- or supercritical depending on the
precise value of the delay parameter $\tau$. A relevant work that
followed the work of Yeung and Strogatz is Ref.~\cite{Montbrio:2006}, which obtained the regions of parameter space corresponding to
synchronized and incoherent solutions, but for particular frequency
distributions. The complex dynamical scenario did
not allow a straightforward analytical treatment to answer the following obvious questions that may be raised for {\it generic frequency distributions}: Can one predict analytically as a function of the time
delay the nature of bifurcation of the incoherent state that one would
observe on changing the value of the coupling constant from low to high
values? What is the critical value of the coupling constant at which the
incoherent state loses its stability? What are the effects of the phase
frustration parameter $\alpha$? Starting from the aforementioned developments and with an aim
to answer the questions just raised, in this work, we embark on a detailed
analytical characterization of bifurcation in model~(\ref{eq:eom}). 

In this work, we will consider ${\cal G}(\omega)$ to be a distribution
 symmetric about its center given by
$\omega=\omega_0$. The quantity $\omega_0$ coincides
with the mean of ${\cal G}(\omega)$ for cases in which the latter
exists. Note that unlike~(\ref{eq:eom-kuramoto}), the
dynamics~(\ref{eq:eom}) is not invariant under the transformation $\theta_j(t)
\to \theta_j(t) - \omega_0 t,~\omega_j \to \omega_j - \omega_0~\forall~j$ that tantamounts
to viewing the dynamics in a frame rotating uniformly at
frequency $\omega_0$ with respect to an inertial frame. From
Eq.~(\ref{eq:eom}), it is evident that viewing the dynamics in such a frame is equivalent to replacing
$\alpha$ with $\alpha'\equiv
\alpha-\omega_0 \tau$. From now on, we will consider such a choice of
the reference frame and consequently the dynamics
\begin{eqnarray}
\dfrac{{\rm d}\theta_j(t)}{{\rm d}t}&=&\omega_j+\dfrac{K}{N}\sum_{k=1}^N
\sin\left[\theta_k(t-\tau)-\theta_j(t)-\alpha+\omega_0 \tau\right]
\nonumber \\ 
&=&\omega_j+K\Im \left[r(t-\tau)e^{-i\left(\theta_j(t)+\alpha-\omega_0
\tau\right)}\right],
\label{eq:eom-1}
\end{eqnarray}
where the $\omega_j$'s are to be regarded as distributed according to
a distribution $g(\omega)$ that is centered at zero: $g(\omega)\equiv
{\cal G}(\omega-\omega_0)$.

The dynamics~(\ref{eq:eom-1}) in the limit $N \to \infty$ is described
by a kinetic equation for the time evolution of the single-oscillator
distribution function $F(\theta,\omega,t)$ that counts the fraction of
oscillators with natural frequency $\omega$ that have their phase
equal to $\theta$ at time $t$. The kinetic equation turns out
to be an infinite-dimensional delay differential equation (DDE), which
has the incoherent state as its stationary solution existing for all
values of $K$. In order to answer our queries raised above, namely, the
bifurcation from the incoherent state that occurs as $K$ is increased to
high values, as a first step, we specialize to a Lorentzian distribution
for the frequencies, and employ the celebrated Ott-Antonsen (OA) ansatz
to deduce from the kinetic equation a simpler DDE satisfied by the order
parameter $r(t)$. 

The Ott-Antonsen (OA) approach offers a powerful exact method to study
the dynamics of coupled
oscillator ensembles~\cite{Ott:2008,Ott:2009}. The approach allows to rewrite in the
thermodynamic limit the dynamics of coupled networks of phase
oscillators in terms of a few collective variables. In the
context of the Kuramoto model~(\ref{eq:eom-kuramoto}) with a Lorentzian distribution of the
oscillator frequencies, the ansatz studies the evolution in phase space
by considering in the space 
${\cal D}$ of all possible single-oscillator distribution functions $F(\theta,\omega,t)$ a
particular class defined on and remaining confined to a manifold
${\cal M}$ in ${\cal D}$ under the time evolution of the phases. The OA 
ansatz obtains for this particular class of $F(\theta,\omega,t)$ a single first-order ordinary differential equation for the
evolution of the synchronization order parameter $r(t)$. The most
remarkable result of the approach, which explains its power and its
widespread applicability in studying oscillator ensembles, is its ability to capture precisely and
quantitatively through this single equation all, and not just
some, of the order parameter attractors and bifurcations of the Kuramoto
dynamics (which may be obtained either by performing numerical
integration of the $N$ coupled non-linear
equations~(\ref{eq:eom-kuramoto}) for $N \gg 1$ and evaluating $r(t)$ in
numerics, or, by simulating the kinetic system
\cite{Acebron:2005,Balmforth:2000,Carrillo:2018}), for a Lorentzian $g(\omega)$. The ansatz has since its
proposition been successfully applied to a variety of setups involving coupled
oscillators. A few recent contributions are Refs.~\cite{Wolfrum:2016,Pazo:2016,Martens:2016,Laing:2016,Ott:2017,Goldobin:2017,Zhang:2017}.
 
Starting with the derived OA-ansatz-reduced dynamics for the order
parameter, we perform both a
linear and a nonlinear stability analysis of the
incoherent state, based on a general formalism to treat
DDE~\cite{Hale:1963,Hale:1993}. 
The linear analysis locates the critical threshold $K_c$ above
which a synchronized state bifurcates from the linearly-unstable incoherent state. On the other hand, the
nonlinear analysis, developed in the spirit of
Refs.~\cite{Crawford:1994,Crawford:1994-1,Crawford:1995,Crawford:1995-1,Crawford:1999,Barre:2016}, 
treats the dynamical flow on the unstable manifold passing through
$K_c$, thereby obtaining the amplitude dynamics of the linearly unstable modes for
$K>K_c$. The form of amplitude dynamics describing the flow in the regime of weak linear
instability, namely, as $K \to K_c^+$, allows to directly obtain the
nature of bifurcation from the incoherent state occurring as soon as $K$
is increased beyond $K_c$. In the second part of our analysis, we relax
the choice of a Lorentzian $g(\omega)$ and perform the linear and
nonlinear stability analysis directly on the kinetic equation for
$F(\theta,\omega,t)$ to derive the amplitude dynamics of the linearly
unstable modes close to the bifurcation of the incoherent state, that
is, as $K \to K_c^+$. Remarkably, we find that the amplitude
equation derived in the general case has the same form as that obtained
from the OA-ansatz-reduced dynamics derived for Lorentzian
$g(\omega)$, thus confirming the power and general applicability of the
OA ansatz. Viewing the analysis of the OA-ansatz-reduced dynamics
\textit{vis-\`{a}-vis} that of the kinetic
equation, we may comment that the former is after all
based on an ansatz that works only for Lorentzian $g(\omega)$, but allows to derive an amplitude dynamics that
works for any $K > K_c$. On the other hand, the analysis based on the
kinetic equation is more general in the sense that there is no ansatz and no specific choice of $g(\omega)$ involved,
but a disadvantage is that the derived amplitude dynamics is valid only close to
the bifurcation, namely, for $K \to K_c^+$.
Quite interestingly and remarkably, we find that close to the
bifurcation, the unstable manifold derived from the kinetic equation has the same form as the OA manifold, so we may say that the OA-ansatz
form may be obtained also as a result of the unstable manifold expansion.

Furthermore, we demonstrate with our exact results a remarkable effect
of time delay: Considering a sum of two Lorentzians as a representative
example of a bimodal distribution, it is known that in the absence of
delay, the bifurcation of the synchronized from the incoherent state is
subcritical~\cite{Martens:2009}. We show here that presence of even a small amount of
delay suffices to completely change the nature of the bifurcation and
make it supercritical ! 

The paper is organized as follows. In Section~\ref{sec:kinetic-equation},
we discuss the characterization of the dynamics~(\ref{eq:eom-1}) in
the thermodynamic limit in terms of a kinetic
equation for the
single-oscillator distribution function $F(\theta,\omega,t)$. In
Section~\ref{sec:Ott-Antonsen}, we apply for the special case of a
Lorentzian distribution for the natural frequencies the Ott-Antonsen ansatz to replace the kinetic equation
with an equation for the order parameter. The linear and nonlinear
stability analysis of the incoherent state is then pursued in detail in
Section~\ref{sec:application-to-OA}, based on a formalism for
treating DDE summarized in Section~\ref{sec:delayDE}. A similar analysis
for the case of general
frequency distributions and applied directly to the kinetic equation is taken up in Section~\ref{sec:application-to-kinetic-equation}, obtaining results
similar to those in Section~\ref{sec:application-to-OA} for Lorentzian
$g(\omega)$. The paper ends with
conclusions and perspectives.

%%%%%%%%%%%%%%%%%%%%%%%%%%%%%%%%%%%%%%%%%%%%%%%%%%%%%%%%%%%%%%%%%%%%%%%%%%%%%%%%%%%%%%
\section{Description of the dynamics in terms of kinetic equation}
\label{sec:kinetic-equation}

In the limit $N \to \infty$, the state of the oscillator
system~(\ref{eq:eom-1}) at time
$t$ may be characterized by the time-dependent single-oscillator
distribution function $F(\theta,\omega,t)$, defined such that $F(\theta,\omega,t){\rm
d}\theta$ gives the fraction of oscillators with frequency $\omega$ that
have their phase in $[\theta,\theta+{\rm d}\theta]$ at time $t$. The
function $F(\theta,\omega,t)$ is $2\pi$-periodic in $\theta$, i.e.,
$F(\theta+2\pi,\omega,t)=F(\theta,\omega,t)$, and is normalized as 
\begin{equation}
\int_0^{2\pi}{\rm d}\theta~F(\theta,\omega,t)=g(\omega)~\forall~t.
\label{eq:f-normalization}
\end{equation}
Since the dynamics~(\ref{eq:eom-1}) conserves in time the number of
oscillators with a given natural frequency, the function  
$F(\theta,\omega,t)$ evolves in time according to a kinetic equation
given by the continuity
equation $\partial F/\partial t+(\partial /\partial \theta)(F{\rm
d}\theta/{\rm d}t)=0$, which on using Eq.~(\ref{eq:eom-1}) yields
(for a rigorous derivation, see Ref.~\cite{Frank:2007})
\begin{eqnarray}
&&\dfrac{\partial F(\theta,\omega,t)}{\partial t}+\omega\dfrac{\partial
F(\theta,\omega,t)}{\partial
\theta}+\dfrac{K}{2i}\dfrac{\partial }{\partial
\theta}\Big[\Big(r[F](t-\tau)e^{-i\left(\theta+\alpha-\omega_0
\tau\right)}\nonumber \\
&&\qquad -r^\ast[F](t-\tau)e^{i\left(\theta+\alpha-\omega_0
\tau\right)}\Big
)F(\theta,\omega,t)\Big]=0.
\label{eq:kinetic-equation}
\end{eqnarray}
Here and in the following, $\ast$ stands for complex conjugation, and we
have defined as functionals of $F$ the quantity
\begin{equation}
r[F](t) \equiv \int {\rm d}\theta {\rm d}\omega~e^{i\theta}F(\theta,\omega,t)
\label{eq:rf-definition}
\end{equation}
as the thermodynamic limit of Eq.~(\ref{eq:rt-definition}).

Equation~(\ref{eq:kinetic-equation}) is an infinite dimensional DDE. In the following section, we
employ the so-called Ott-Antonsen ansatz~\cite{Ott:2008,Ott:2009} that
allows to derive from Eq.~(\ref{eq:kinetic-equation}) a DDE for the order parameter.

%%%%%%%%%%%%%%%%%%%%%%%%%%%%%%%%%%%%%%%%%%%%%%%%%%%%%%%%%%%%%%%%%%%%%%%%%%%%%%%%%%%%%%
\section{The Ott-Antonsen-ansatz-reduced dynamics}
\label{sec:Ott-Antonsen}

The implementation of the Ott-Antonsen ansatz for the
dynamics~(\ref{eq:eom-1}) has been
discussed in Ref.~\cite{Ott:2008}, which we briefly recall here for use
in later parts of the paper. As is usual with OA ansatz
implementation, we will make the specific choice of a Lorentzian
distribution for $g(\omega)$:
\begin{equation}
g(\omega)=\dfrac{\Delta}{\pi}\dfrac{1}{\omega^2+\Delta^2},
\label{eq:lorentzian-gw}
\end{equation}
where $\Delta >0$ denotes the half-width-at-half-maximum of $g(\omega)$. 
Consider the function  
$F(\theta,\omega,t)$, which being $2\pi$-periodic in $\theta$ may be
expanded in a Fourier series in $\theta$:
\begin{eqnarray}
&&F(\theta,\omega,t)=\dfrac{g(\omega)}{2\pi}\Big[1+\sum_{n=1}^\infty
\Big(\widetilde{F}_n(\omega,t)e^{in\theta}+[\widetilde{F}_n(\omega,t)]^\ast
e^{-in\theta}\Big)\Big],
\label{eq:f-expansion}
\end{eqnarray}
where $\widetilde{F}_n(\omega,t)$ is the $n$-th Fourier coefficient. Using $\int_0^{2\pi}{\rm
d}\theta~e^{in\theta}=2\pi \delta_{n,0}$, we check that the above
expansion is consistent with Eq.~(\ref{eq:f-normalization}).

The OA ansatz considers in the expansion~(\ref{eq:f-expansion}) a restricted class of Fourier coefficients given
by~\cite{Ott:2008,Ott:2009}
\begin{equation}
\widetilde{F}_n(\omega,t)=[z(\omega,t)]^n,
\label{eq:OA}
\end{equation}
with $z(\omega,t)$ an arbitrary function with the restriction
$|z(\omega,t)|<1$ that makes the infinite series in
Eq.~(\ref{eq:f-expansion}) a convergent one. In implementing the OA
ansatz, it is also assumed that $z(\omega,t)$ may be
analytically continued to the whole of the complex-$\omega$ plane, that it has no singularities in the lower-half complex-$\omega$ plane, and that
$|z(\omega,t)|\to 0$ as $\Im(\omega) \to -\infty$~\cite{Ott:2008,Ott:2009}. 
Using Eqs.~(\ref{eq:f-expansion}) and~(\ref{eq:OA}) in
Eq.~(\ref{eq:rf-definition}), one gets
\begin{equation}
r(t)\equiv r[F](t)=\int_{-\infty}^\infty {\rm
d}\omega~g(\omega)z^\ast(\omega,t). 
\label{eq:rf-OA}
\end{equation}
On substituting Eqs.~(\ref{eq:f-expansion}),~(\ref{eq:OA}),
and~(\ref{eq:rf-OA}) in Eq.~(\ref{eq:kinetic-equation}) and on collecting and equating the
coefficient of $e^{in\theta}$ to zero, we get 
\begin{eqnarray}
&&\dfrac{\partial z(\omega,t)}{\partial t}+i\omega
z(\omega,t)+\dfrac{K}{2}\Big[e^{-i\left(\alpha-\omega_0 \tau\right)}
r(t-\tau)z^2(\omega,t)-e^{i\left(\alpha-\omega_0
\tau\right)}r^\ast(t-\tau)\Big]=0. \nonumber \\
\label{eq:alpha-differential-equation}
\end{eqnarray}
Note that even if the initial condition does not respect the OA ansatz, it has been shown that the OA manifold is attractive for the dynamics~\eqref{eq:kinetic-equation} with $\tau=0$, (see,  for example, Ref.~\cite{dietert_asymp}).

For the Lorentzian $g(\omega)$, Eq.~(\ref{eq:lorentzian-gw}), one may
evaluate $r(t)$ by using Eq.~(\ref{eq:rf-OA}) to get
\begin{eqnarray}
r(t)&=&\dfrac{1}{2i\pi}\oint_C {\rm
d}\omega~z^\ast(\omega,t)\left[\dfrac{1}{\omega-i\Delta}-\dfrac{1}{\omega+i\Delta}\right]\nonumber
\\
&=&z^\ast(-i\Delta,t),
\label{eq:rf-integral}
\end{eqnarray}
where the contour $C$ consists of the real-$\omega$ axis closed by a
large semicircle in the lower-half complex-$\omega$ plane on which the
integral in Eq.~(\ref{eq:rf-integral}) gives zero contribution in view of
$|z(\omega,t)|\to 0$ as $\Im(\omega) \to -\infty$. The second
equality in Eq.~(\ref{eq:rf-integral}) is obtained by applying the
residue theorem to evaluate the complex integral over the contour $C$.
Using Eqs.~(\ref{eq:alpha-differential-equation})
and~(\ref{eq:rf-integral}), we finally obtain the OA equation for the time
evolution of the synchronization order parameter as the DDE~\cite{Ott:2008}
\begin{eqnarray}
&&\dfrac{{\rm d} r(t)}{{\rm d}
t}+\Delta r(t)+\dfrac{K}{2}\Big[e^{-i\left(\alpha-\omega_0
\tau\right)}r^\ast(t-\tau)r^2(t)-e^{i\left(\alpha-\omega_0\tau\right)}r(t-\tau)\Big]=0,
\label{eq:rt-differential-equation}
\end{eqnarray}
whose solution requires as an initial condition the value of $r(t)$ over an entire interval of time $t$, namely, $t \in
[-\tau,0]$.
Note that for $\tau=0$, Eq.~(\ref{eq:rt-differential-equation}) is a
finite-dimensional ODE for $r(t)$ that requires for its solution only
the value of $r(t)$ at $t=0$ as an initial condition. In this case, it has been demonstrated that
this single equation contains all the
bifurcations and attractors of the order
parameter~(\ref{eq:rt-definition}) as obtained through the evolution of
the set of coupled differential equations~(\ref{eq:eom-kuramoto}) for a
Lorentzian $g(\omega)$ and in the limit $N \to \infty$. 
In order to effectively describe the bifurcation of the
dynamics~(\ref{eq:rt-differential-equation}) between an incoherent and a
synchronized state in the case $\tau \ne 0$, we need to put the
system~(\ref{eq:rt-differential-equation}) into the so-called normal form,
by reducing the dimensionality of the
evolution~(\ref{eq:rt-differential-equation}) to get a simple ODE. We
take up this program in Section~\ref{sec:application-to-OA} based on
a general formalism for DDE that we summarize in the next section.

%%%%%%%%%%%%%%%%%%%%%%%%%%%%%%%%%%%%%%%%%%%%%%%%%%%%%%%%%%%%%%%%%%%%%%%%%%%%%%%%%%%%%%
\section{Theory of delay differential equation}
\label{sec:delayDE}

It is evident from the form of equations~(\ref{eq:kinetic-equation})
and~(\ref{eq:rt-differential-equation}) that both may be cast in the general
form
\begin{equation}
\dfrac{\partial}{\partial t}H(t)=\mathscr{M}[H(t),H_t(-\tau)],
\label{eq:gt-differential-equation}
\end{equation}
where the function $H(t)$ is either the function $F(\theta,\omega,t)$ or
the function $r(t)$ depending on the case of interest. Here, we have introduced the notation
$H_t(\varphi)\equiv H(t+\varphi)$ for
$-\tau\leq \varphi\leq 0$. To solve for $H(t);~t> 0$ using
Eq.~(\ref{eq:gt-differential-equation}), one must
specify as an initial
condition the function $H_t(\varphi);~-\tau \le \varphi \le 0$. The time
evolution of $H_t(\varphi)$ for $-\tau \le \varphi <0$ is obtained as
$\partial H_t(\varphi)/\partial
t=\lim_{\delta\to 0}(H_{t+\delta}(\varphi)-H_t(\varphi))/\delta=\lim_{\delta \to
0}(H(t+\varphi+\delta)-H(t+\varphi))/\delta={\rm d}H(t+\varphi)/{\rm d}\varphi={\rm
d}H_t(\varphi)/{\rm d} \varphi$. We may then quite generally write
\begin{equation}
\dfrac{\partial}{\partial t}H_t(\varphi)=(\mathscr{A}
H_t)(\varphi);~~-\tau\leq \varphi\leq 0,
\label{eq:DE-general-form}
\end{equation}
where we have
\begin{equation}
(\mathscr{A} H_t)(\varphi)=
\begin{cases}
\dfrac{\mathrm d}{\mathrm d\varphi}H_t(\varphi);~~-\tau\leq\varphi< 0,\\
\mathscr{M}[H_t(\varphi),H_{t-\tau}(\varphi)];~~\varphi=0.
\end{cases}
\end{equation}

Around a stationary state $H_{\rm{st}}$ of
Eq.~(\ref{eq:DE-general-form}), we define the perturbation
$h_t(\varphi)$ as $H_t(\varphi)\equiv H_{\rm{st}}+h_t(\varphi)$. The
time evolution of $h_t(\varphi)$ has the same form as
Eq.~(\ref{eq:DE-general-form}), in which we now split the operator $\mathscr{A}$ into a part $\mathscr{D}$ that is
linear in $h$ and a part $\mathscr{F}$ that is nonlinear, to write
\begin{eqnarray}
&&(\mathscr{A} h_t)(\varphi)=(\mathscr{D}h_t+\mathscr{F}[h_t]
)(\varphi)\nonumber \\
&&=\begin{cases}
\dfrac{\mathrm d}{\mathrm d\varphi}h_t(\varphi)\\
\mathscr{L}h_t(\varphi)
\end{cases}
+
\begin{cases}
0;~~-\tau\leq\varphi< 0,\\
\mathscr{N}[h_t];~~\varphi=0.
\end{cases}
\end{eqnarray}
We will find it convenient to decompose the linear operator
$\mathscr{L}$ into a part $L$ that does not contain any term involving the delay and a part
$R$ containing the delay terms:
$\mathscr{L}h_t(\varphi)=\operatorname{L}h_t(0)+\operatorname{R}h_t(-\tau)$.

The adjoint of the linear operator $\mathscr{D}$ may be defined through the
definition of the scalar product~\cite{Hale:1963}
\begin{eqnarray}
&&(h_{1t},h_{2t})_\tau\equiv
(h_{1t}(0),h_{2t}(0))
+\int_{-\tau}^0 {\mathrm d}\xi~ \left(h_{1t}(\xi+\tau),\operatorname{R} h_{2t}(\xi)\right),
\label{eq:definition-scalar-product}
\end{eqnarray}
where the scalar product in the absence of
any delay $(\cdot,\cdot)$ is either 
\begin{equation}
\left( s,r\right)=s^\ast r,\quad\text{with}~ h_1(0)=s,~ h_2(0)=r
\end{equation}
for the finite dimensional case, or, the standard $\mathcal{L}_2(\mathbb{T}\times\mathbb{R})$ scalar product
\begin{equation}
\left( h,f\right)=\int_{\mathbb{T}\times\mathbb{R}}h^\ast(\theta,\omega) f(\theta,\omega)\,\mathrm{d}\omega\,\mathrm{d} \theta ,\quad \text{with}~ h_1(0)=h(\theta,\omega),~h_2(0)=f(\theta,\omega), 
\end{equation}
for the kinetic case (infinite dimensional).
The adjoint operator $\mathscr{D}^\dagger$, which satisfies
\begin{equation}
(h_{1t}(\varphi),\mathscr{D}h_{2t}(\varphi))_\tau=(\mathscr{D}^\dagger h_{1t}(\varphi),
h_{2t}(\varphi))_\tau,
\end{equation}
is obtained as
\begin{eqnarray}
(\mathscr{D}^\dagger h_t)(\vartheta)=\begin{cases}
-\dfrac{\mathrm d}{{\mathrm d}\vartheta}h_t(\vartheta);~~0<\vartheta\leq\tau, \\
\mathscr{L}^\dagger h_t(\vartheta)=\operatorname{L}^\dagger h_t(0)+\operatorname{R}^\dagger
h_t(\tau);~~\vartheta=0.
\end{cases}
\end{eqnarray}
Note that $h_2(\varphi)$ and $h_1(\vartheta)$ belong to different functional spaces, for example, $\varphi\in [-\tau,0]$ and $\vartheta\in [0,\tau]$. 
%%%%%%%%%%%%%%%%%%%%%%%%%%%%%%%%%%%%%%%%%%%%%%%%%%%%%%%%%%%%%%%%%%%%%%%%%%%%%%%%%%%%%%
\section{Application to the Ott-Antonsen-reduced dynamics}
\label{sec:application-to-OA}

We now apply the formalism of the DDE discussed in the preceding section
to analyze Eq.~(\ref{eq:rt-differential-equation}), with the aim to reduce this
infinite-dimensional equation to a finite-dimensional equation. 
The incoherent stationary state $r_{\rm st}=0$ is always a solution
of Eq.~(\ref{eq:rt-differential-equation}). It is of interest to study its
stability as $K$ is varied,
which may be done by studying the behavior of perturbations about
$r_{\rm st}=0$, where the perturbations $r_t(\varphi)$ satisfy
\begin{equation}
\dfrac{\mathrm d}{\mathrm d t} r_t(\varphi)=(\mathscr{D}
r_t+\mathscr{F}[r_t])(\varphi);~~-\tau \le \varphi \le 0,
\label{eq:rt-differential-equation-1}
\end{equation}
with 
\begin{eqnarray}
(\mathscr{D}r_t)(\varphi)=\begin{cases}
\dfrac{\mathrm d}{\mathrm d\varphi}r_t(\varphi);~~- \tau\leq\varphi<0,\\
\mathscr{L}r_t(\varphi)=\operatorname{L} r_t(0)+\operatorname{R} r_t(-\tau);~~\varphi=0,
\end{cases}
\end{eqnarray}
and 
\begin{equation}
(\mathscr{F}[r_t])(\varphi)=\begin{cases}
0;~~- \tau\leq\varphi<0,\\
\mathscr{N}[r_t];\quad \varphi=0.
\end{cases}
\end{equation}
The adjoint of the linear operator $\mathscr{D}$ is given by
\begin{eqnarray}
(\mathscr{D}^\dagger s_t)(\vartheta)=\begin{cases}
-\dfrac{\mathrm d}{\mathrm d\vartheta}s_t(\vartheta);~~0<\vartheta\leq \tau,\\
\mathscr{L}^\dagger s_t(\vartheta)=\operatorname{L}^\dagger s_t(0)+\operatorname{R}^\dagger
s_t(\tau);~~\vartheta=0.
\end{cases}
\end{eqnarray}
In these equations, we have
\begin{eqnarray}
&& \operatorname{L} r=-\Delta r,\\
&&\operatorname{R} r=\dfrac{K}{2} e^{i(\alpha-\omega_0\tau)}r, \\
&&\operatorname{L}^\dagger r=-\Delta r,\\
&&\operatorname{R}^\dagger r=\dfrac{K}{2}e^{-i(\alpha-\omega_0\tau)}r, \\
&&\mathscr{N}[r_t]=-\dfrac{K}{2}e^{-i(\alpha-\omega_0\tau)}r^\ast_t(-\tau) r_t^2(0).
\end{eqnarray}

Small perturbations $r_t(\varphi)$ may be
expressed as a linear combination of the eigenfunctions of the linear
operator $\mathscr{D}$. To this end, let us then solve the eigenfunction equation
\begin{equation}
(\mathscr{D}p)(\varphi)=\lambda p(\varphi)
\end{equation}
for $- \tau\leq\varphi<0$; we get $p(\varphi)=p(0) e^{\lambda\varphi}$.
The equation $(\mathscr{D}p)(\varphi)=\lambda p(\varphi)$ for
$\varphi=0$ gives $\lambda p(0)=-\Delta
p(0)+(K/2)e^{i(\alpha-\omega_0\tau)}p(0)e^{-\lambda \tau}$. With
$p(0)\ne 0$, we thus get the dispersion relation
\begin{equation}
\Lambda(\lambda)=\lambda+\Delta-\dfrac{K}{2}e^{-\lambda\tau+i(\alpha-\omega_0\tau)}=0.
\label{eq:OA-dispersion}
\end{equation}
The solution of the above equation gives the discrete eigenvalues
$\lambda_l$ (with $l\in\mathbb{Z}$) in terms of the Lambert-W function
$W_l$, as
\begin{equation}
\lambda_l=\dfrac{-\Delta  \tau +W_l\left(\dfrac{K \tau}{2} e^{i\alpha
+\Delta\tau -i\omega_0\tau}\right)}{\tau}.
\end{equation}
Without loss of generality, we may take $p(0)=1$. We thus conclude that
$p(\varphi)=e^{\lambda \varphi}$ is an eigenfunction
of the linear operator $\mathscr{D}$ for $-\tau \le \varphi \le 0$ with
eigenvalue $\lambda$ provided that $\lambda$ satisfies $\Lambda(\lambda)=0$. In other words, a discrete set of
eigenvalues correspond to $\mathscr{D}$ for all values of $\varphi$.
Perturbations $r_t(\varphi)$ may be
expressed as a linear combination of the corresponding eigenfunctions. It then
follows that the stationary solution $r_{\rm st}=0$
will be linearly stable under the
dynamics~(\ref{eq:rt-differential-equation-1}) so long as all the eigenvalues
$\lambda$ have a real part that is negative. Vanishing of the real part
of the eigenvalue with the smallest real part then signals criticality above which
$r_{\rm st}=0$ is no longer a linearly-stable stationary solution of
Eq.~(\ref{eq:rt-differential-equation-1}). Denoting by $\lambda_{\rm
i};~\lambda_{\rm i}\in \mathbb{R}$ the
imaginary part of the eigenvalue with the smallest real part, we thus have at
criticality the following equations obtained from
Eq.~(\ref{eq:OA-dispersion}):
\begin{eqnarray}
&&\dfrac{K_c}{2}\cos(\alpha-(\omega_0+\lambda_{\rm i})\tau)=\Delta,
\nonumber \\
\label{eq:Kc-equations-lorentzian} \\
&&\dfrac{K_c}{2}\sin(\alpha-(\omega_0+\lambda_{\rm
i})\tau)=\lambda_{\rm i}. \nonumber
\end{eqnarray}

We want to study the behavior of $r_t(\varphi)$ as $K \to K_c^+$, the goal being
to uncover the weakly nonlinear dynamics occurring beyond the
exponential growth taking place due to the instability as $K \to K_c^+$. To this end,
we want to study the behavior of $r_t(\varphi)$ on the unstable
manifold, which by definition is tangential to the unstable eigenspace
spanned by the eigenfunctions $p(\varphi)$ at the equilibrium point
$(K=K_c,\lambda=i\lambda_{\rm i})$. This manifold may be shown to be an
attractor of the dynamics for the type of DDE under consideration~\cite{Hale:1993,Murdock:2006,Guo:2013}
and is therefore of interest to study. To proceed, we need
the eigenfunctions of the adjoint operator $\mathscr{D}^\dagger$, which
will be useful in discussing the unstable manifold expansion. It is
easily checked that $\mathscr{D}^\dagger$ has the eigenfunction 
$q(\vartheta)=q(0)e^{-\lambda^\ast\vartheta}$ associated with the eigenvalue
$\lambda^\ast$
satisfying $\Lambda^\ast(\lambda^\ast)=0$, that is, we get the same dispersion
relation as for $\mathscr{D}$. We may choose $q(0)$ such that
$(q(\varphi),p(\varphi))_\tau=1$. Using
Eq.~(\ref{eq:definition-scalar-product}), and
noting that in the present case, $(q(0),p(0))=q^\ast(0)p(0)$, we get
\begin{equation}
q^\ast(0)+\int_{-\tau}^0
{\mathrm d}\xi~q^\ast(0)e^{-\lambda(\xi+\tau)}\dfrac{K}{2}e^{i(\alpha-\omega_0)\tau}e^{\lambda
\xi}=1,
\end{equation}
yielding
\begin{equation}
q^\ast(0)=\dfrac{1}{1+\tau\dfrac{K}{2}e^{-(\lambda+i\omega_0)\tau+i\alpha}}=\dfrac{1}{\Lambda^\prime(\lambda)}
\label{eq:OA-qstar-equation}
\end{equation}
where we chose without loss of generality $p(0)=1$.

The unstable
manifold expansion of $r_t(\varphi)$ for $K>K_c$ reads
\begin{equation}
r_t(\varphi)=A(t)p(\varphi)+w[A](\varphi),
\label{eq:rt-expansion}
\end{equation}
where $w[A](\varphi)$, which is at least quadratic in $A$ (in fact, one
can prove that it is cubic in $A$ in the present case), denotes the component of $r_t(\varphi)$ transverse
to the unstable eigenspace, so that $(q(\varphi),w(\varphi))_\tau=0$.
On using the latter equation, together with
$(q(\varphi),p(\varphi))_\tau=1$ in Eq.~(\ref{eq:rt-expansion}), we get
$A(t)=(q(\varphi),r_t(\varphi))_\tau$. 
The time evolution of $A(t)$ is then obtained as 
\begin{eqnarray}
\dot{A}&=&(q(\varphi),\dot{r}_t(\varphi))_\tau\nonumber \\
&=&(q(\varphi),(\mathscr{D}r_t+\mathscr{F}[r_t])(\varphi))_\tau\nonumber\\
&=&(q(\varphi),A(t)\lambda
p(\varphi)+\mathscr{D}w(\varphi)+\mathscr{F}[r_t](\varphi))_\tau\nonumber
\\&=&\lambda A+q^\ast(0)\mathscr{N}[r_t],
\label{eq:OA-A-equation}
\end{eqnarray}
where the dot denotes derivative with respect to time.
Here, in arriving at the second and the third equality, we have used
Eqs.~(\ref{eq:rt-differential-equation-1}) and (\ref{eq:rt-expansion}),
while in obtaining the last equality, we have used in the third step 
$(q(\varphi),\mathscr{D}w(\varphi))_\tau=(\mathscr{D}^\dagger
q(\varphi),w(\varphi))_\tau=\lambda^\ast(q(\varphi),w(\varphi))_\tau=0$.
Since we can prove that $w$ is $O(|A|^2A)$, while we see that $\mathscr{N}[r_t]$ is of order three in
$r_t$, the leading-order contribution to the nonlinear term on the
right hand side of Eq.~(\ref{eq:OA-A-equation}) is obtained
as $\mathscr{N}[r_t]= \mathscr{N}[A(t)p(\varphi)]+O(|A|^2A)=-(K/2)e^{-i(\alpha-\omega_0\tau)}p^\ast(-\tau)p^2(0)|A(t)|^2 A(t)]+O(|A|^2A)$. Using
this result and Eq.~(\ref{eq:OA-qstar-equation}) in
Eq.~(\ref{eq:OA-A-equation}), we get eventually the so-called normal form for the time evolution of
$A$ as
\begin{equation}
\dot{A}=\lambda
A-\dfrac{K}{2}\dfrac{e^{(i\omega_0-\lambda^\ast)\tau-i\alpha}}{1+\tau
\dfrac{K}{2}e^{-(\lambda+i\omega_0)\tau+i\alpha}}|A|^2 A+O(|A|^2A).
\label{eq:OA-normal-form}
\end{equation}
The above is the desired finite-dimensional ordinary differential equation
corresponding to the infinite-dimensional equation
(\ref{eq:rt-differential-equation-1}), which allows to decide the
bifurcation behavior of $r_t(\varphi)$ as $K \to K_c^+$. The relevant
parameter to study the type of bifurcation is given by the sign of the
second term on the right hand side. Denoting this term by $c_3$, we then need to
study the sign of the real part of $c_3$ as the real part of $\lambda$
approaches zero, so that $\lambda=i\lambda_{\rm i}$ is purely imaginary:
\begin{equation}
\Re(c_3)=-\dfrac{K_c}{2}\Re\left
(\dfrac{e^{i(\omega_0+\lambda_{\rm i})\tau-i\alpha}}{1+\tau
\dfrac{K_c}{2}e^{-i(\omega_0+\lambda_{\rm i})\tau+i\alpha}}\right ).
\label{eq:OA-c3}
\end{equation}
Note that a similar approach was pursued in \cite{yguo}.
%%%%%%%%%%%%%%%%%%%%%%%%%%%%%%%%%%%%%%%%%%%%%%%%%%%%%%%%%%%%%%%%%%%%%%%%%%%%%%%%%%%%%%
\section{Application to the kinetic equation}
\label{sec:application-to-kinetic-equation}
It is of interest to consider the formalism of the DDE and the unstable manifold
expansion applied in the previous section to the OA-ansatz-reduced
dynamics~(\ref{eq:rt-differential-equation}), and to apply it directly to the kinetic system,
Eq.~(\ref{eq:kinetic-equation}), so as to reduce this
infinite-dimensional equation to a finite-dimensional one. The advantage
of such an application stems from the fact that while the OA-ansatz-reduced-dynamics is obtained for a Lorentzian
$g(\omega)$, the kinetic system is valid for any generic $g(\omega)$. A
disadvantage is that the computations for the kinetic system are only formal in the
sense that standard theorems proving the attracting property of the unstable
manifold do not apply straightforwardly for the type of kinetic equation
under consideration because of the existence of the continuous spectrum
(see below). y attractiveness of a manifold is meant that almost all trajectories during the course of the dynamical evolution come at long times arbitrarily close to the manifold so that their eventual evolution coincides with evolution of trajectories lying on the manifold itself. The reader is referred to Refs. \cite{Chiba:2013,Dietert:2016} for a mathematical proof of the statement in the non-delayed model, that, is, for $\tau=0$.

Let us give at the outset an outline of the current rather technical section. We start off with rewriting of the kinetic equation~(\ref{eq:kinetic-equation}), of which the incoherent state $F_{\rm st}$ represents a stationary solution, in the form of a DDE for perturbations $f_t(\varphi)$ around $F_{\rm st}$. The DDE involves a linear evolution operator $\mathscr{D}$ and a nonlinear one. In the next step, we obtain the eigenvalues and the eigenvectors of $\mathscr{D}$ and of the corresponding adjoint operator $\mathscr{D}^\dagger$. As is usual with linear stability analysis~\cite{Strogatz-book}, the knowledge of the eigenvalues allows to locate the critical value $K_c$ of the coupling $K$ above which the incoherent state $F_{\rm st}$ becomes linearly unstable and a synchronized stationary state bifurcates from it. Following this, we perform for $K>K_c$ an unstable manifold expansion of $F_{\rm st}$ along the two unstable eigenvectors (representing respectively the eigenvectors of $\mathscr{D}$ and $\mathscr{D}^\dagger$) and the unstable manifold passing through $K_c$. By invoking a convenient Fourier expansion of the relevant quantities and working at $K$ slightly greater than $K_c$, we obtain the amplitude dynamics describing the evolution of perturbations $f_t(\varphi)$ in the regime of weak linear instability, $K\to K_c^+$. The form of the amplitude dynamics allows to directly obtain the nature of bifurcation occurring as soon as $K$ is increased beyond $K_c$: The amplitude dynamics has a leading linear term, followed by a nonlinear (cubic) term, and according to elements of bifurcation theory well known in the literature \cite{Strogatz-book}, it is the sign of this cubic term that dictates the precise nature of the bifurcation (positive and negative leading respectively to subcritical and supercritical bifurcation). We show explicitly how the sign of the cubic term varies as a function of the delay $\tau$, for two separate cases, namely, that of a unimodal and a bimodal Lorentzian frequency distribution. A remarkable result of this section is that close to the bifurcation, the unstable manifold that we derive in this section based on the kinetic equation~(\ref{eq:kinetic-equation}) has the same form as the Ott-Antonsen manifold discussed in Section~\ref{sec:Ott-Antonsen}. We now proceed to a detailed derivation of our results.

Similar to the preceding section, we rewrite
Eq.~(\ref{eq:kinetic-equation}) in the
form of Eq.~(\ref{eq:DE-general-form}). We consider perturbations $f_t(\varphi)$
around the incoherent stationary state of
(\ref{eq:kinetic-equation}), namely, $F_{\rm st}=g(\omega)/(2\pi)$, so
that $F_t(\varphi)=F_{\rm st}(\varphi)+f_t(\varphi)$. From
Eq.~(\ref{eq:f-normalization}),
it follows that $\int_0^{2\pi}{\rm d}\theta~f_t(\varphi)=0$. Perturbations $f_t$
will evolve according to
\begin{equation}
\dfrac{\mathrm d}{\mathrm d t} f_t(\varphi)=(\mathscr{D}
f_t+\mathscr{F}[f_t])(\varphi);~~-\tau \le \varphi \le 0,
\label{eq:ft-differential-equation-1}
\end{equation}
with
\begin{eqnarray}
(\mathscr{D}f_t)(\varphi)=\begin{cases}
\dfrac{\mathrm d}{\mathrm d\varphi}f_t(\varphi);~~- \tau\leq\varphi<0,\\
\mathscr{L}f_t(\varphi)=\operatorname{L} f_t(0)+\operatorname{R} f_t(-\tau);~~\varphi=0,
\end{cases}
\end{eqnarray}
and 
\begin{eqnarray}
(\mathscr{F}[f_t])(\varphi)=\begin{cases}
0;~~- \tau\leq\varphi<0,\\
\mathscr{N}[f_t];\quad \varphi=0.
\end{cases}
\end{eqnarray}
The adjoint of the linear operator $\mathscr{D}$ is given by
\begin{eqnarray}
(\mathscr{D}^\dagger h_t)(\vartheta)=\begin{cases}
-\dfrac{\mathrm d}{\mathrm d\vartheta}h_t(\vartheta);~~0<\vartheta\leq \tau,\\
\mathscr{L}^\dagger h_t(\vartheta)=\operatorname{L}^\dagger h_t(0)+\operatorname{R}^\dagger
h_t(\tau);~~\vartheta=0.
\end{cases}
\end{eqnarray}
Here, we have 
\begin{eqnarray}
&&\operatorname{L} f=-\omega\dfrac{\partial }{\partial \theta}f,\\
&&\operatorname{R} f=\dfrac{K}{2}\dfrac{g(\omega)}{2\pi}\left (r[f]
e^{-i(\theta+\alpha-\omega_0\tau)}+r^\ast[f]e^{i(\theta+\alpha-\omega_0\tau)}\right
), \\
&&\operatorname{L}^\dagger h=\omega\dfrac{\partial }{\partial \theta} h,\\
&&\operatorname{R}^\dagger h=\dfrac{K}{4\pi}\left (r\left [g(\omega)h\right ]
e^{-i(\theta-\alpha+\omega_0\tau)}+r^\ast[g(\omega)h]e^{i(\theta-\alpha+\omega_0\tau)}\right
), 
\\
&&\mathscr{N}[f_t]=-\dfrac{K}{2i}\dfrac{\partial }{\partial \theta}\Big
(\Big(r[f_t](-\tau)e^{-i(\theta+\alpha-\omega_0\tau)}
-r^\ast[f_t](-\tau)e^{i(\theta+\alpha-\omega_0\tau)}\Big)f_t(0)\Big)\nonumber
\\.\label{eq:N}
\end{eqnarray}
To study the linear stability of $F_{\rm st}$,
similar to what was done in the preceding section, we first solve the eigenfunction equation
\begin{equation}
(\mathscr{D}P)(\varphi)=\lambda P(\varphi)
\end{equation}
for $- \tau\leq\varphi<0$; we get $P(\varphi)=\Psi e^{\lambda\varphi}$
for arbitrary $\Psi$. Since we will in the following expand
$f_t(\varphi)$ in terms of $P(\varphi)$, we would need to choose
$\Psi$ as $\Psi(\theta,\omega)$, where $2\pi$-periodicity of $f_t$ implies
that so should be $\Psi(\theta,\omega)$. Consequently, we may expand
$\Psi(\theta,\omega)$ in a Fourier series in $\theta$, as
$\Psi(\theta,\omega)=(2\pi)^{-1}\sum_{k=-\infty}^\infty
\psi_k(\omega)e^{ik\theta}$, so that $P(\varphi)=(2\pi)^{-1}\sum_{k=-\infty}^\infty
\psi_k(\omega)e^{i k\theta}e^{\lambda \varphi}$.
Using the equation $(\mathscr{D}P)(\varphi)=\lambda
P(\varphi)$ for
$\varphi=0$ and $k=\pm 1$ in the Fourier expansion of $P(\varphi)$, it may be easily seen with the condition $r[\Psi]=r^\ast[\Psi]=1$ that
$p(\varphi)=\psi_1(\omega)e^{i\theta+\lambda \varphi}$ and
$p^\ast(\varphi)$ give two independent eigenfunctions of $\mathscr{D}$
with eigenvalues $\lambda$ and $\lambda^\ast$, respectively, where the
latter satisfy $\Lambda(\lambda)=\Lambda^\ast(\lambda^\ast)=0$, and 
\begin{eqnarray}
&&\psi_1(\omega)=\dfrac{K}{2}e^{-\lambda\tau+i(\alpha-\omega_0\tau)}\dfrac{g(\omega)}{\lambda+i\omega},\label{eq:psi_1}
\\
&&\Lambda(\lambda)=1-\dfrac{K}{2}e^{-\lambda\tau+i(\alpha-\omega_0\tau)}\int{\rm
d}\omega~\dfrac{g(\omega)}{\lambda+i\omega}.
\label{eq:dispersion-kinetic-equation}
\end{eqnarray}
For $k\neq\pm 1$, one
 has a continuous spectrum sitting on the imaginary axis; this feature is
characteristic of kinetic equations of the type of
Eq.~(\ref{eq:kinetic-equation})~\cite{Crawford:1989,Crawford:1994,Strogatz:1991}. For $K > K_c$, when the incoherent stationary
state is linearly unstable, the
unstable eigenspace is spanned by the eigenfunctions $p(\varphi)$ and
$p^\ast(\varphi)$.

The stationary solution $F_{\rm st}=g(\omega)/(2\pi)$ will be neutrally
stable, thanks to the continuous spectrum sitting on the imaginary axis
that generates a dynamics similar to that of Landau
damping~\cite{Strogatz:1992,Yeung:1999}, when there are no eigenvalues
$\lambda$ with a positive real part. Vanishing of the real part
of the eigenvalue with the smallest real part then signals criticality above which
$F_{\rm st}$ becomes a linearly unstable stationary solution of
Eq.~(\ref{eq:rt-differential-equation-1}). Denoting by $\lambda_{\rm
i};~\lambda_{\rm i}\in \mathbb{R}$ the
imaginary part of the eigenvalue with the smallest real part, we thus have at
criticality the following equations obtained from
Eq.~(\ref{eq:dispersion-kinetic-equation}):
\begin{eqnarray}
\cos\left (\alpha -(\omega_0+\lambda_{\rm i})\tau \right
)&=&\dfrac{g(-\lambda_{\rm i})\pi K_c}{2}, \nonumber \\
\label{eq:ft-Kc-1} \\
\tan\left (\alpha -(\omega_0+\lambda_{\rm i})\tau \right
)&=&\dfrac{\operatorname{PV}\int\dfrac{g(\omega)}{\omega+\lambda_{\rm
i}}}{\pi
g(-\lambda_{\rm i})}, \nonumber
\end{eqnarray}
where $\operatorname{PV}$ stands for principal value. One then has to have $\cos\left (\alpha -(\omega_0+\lambda_{\rm i})\tau 
\right )>0$ in order to have a solution of the first equation.
For the particular choice of the Lorentzian distribution,
Eq.~(\ref{eq:lorentzian-gw}), Eq.~(\ref{eq:ft-Kc-1}) reduces to
Eq.~(\ref{eq:OA-dispersion}), as it should.

The eigenfunctions of the adjoint operator $\mathscr{D}^\dagger$ are
given by $q(\vartheta)=\widetilde{\psi}_1(\omega)e^{i\theta-\lambda^\ast \vartheta}$ and
$q^\ast(\vartheta)$ with eigenvalues $\lambda^\ast$ and $\lambda$,
respectively, where we fix $\widetilde{\psi}_1(\omega)$ by requiring that
$(q(\varphi),p(\varphi))_\tau=1=\int {\rm d}\theta{\rm
d}\omega~q^\ast(0)p(0)+\int_{-\tau}^0{\rm d}\xi \int {\rm d}\theta
{\rm d}\omega~q^\ast(\xi+\tau)\operatorname{R}p(\xi)$. 
We thus get
$\widetilde{\psi}_1(\omega)=((\Lambda'(\lambda))^\ast(\lambda^\ast-i\omega)2\pi)^{-1}$,
and hence, 
\begin{equation}
q(\vartheta)=\dfrac{1}{(\Lambda'(\lambda))^\ast}\dfrac{1}{\lambda^\ast-i\omega}e^{i\theta-\lambda^\ast\vartheta}.
\end{equation}

Similar to Eq.~(\ref{eq:rt-expansion}), we now decompose perturbations $f_t(\varphi)$
along the two unstable eigenvectors and the unstable manifold, as
\begin{equation}
f_t(\varphi)=A(t)p(\varphi)+A^\ast(t)p^\ast(\varphi)+w[A,A^\ast](\varphi),
\label{eq:decomposition}
\end{equation}
with the relations $(q(\varphi),p(\varphi))_\tau=1,(q(\varphi),p^\ast(\varphi))_\tau=0,(q(\varphi),w(\varphi))_\tau=0$
yielding $A(t)=(q(\varphi),f_t(\varphi))_\tau$. We require $w(\varphi)$
to be at least quadratic in $A$. 

Let us define the following Fourier expansion needed for further
analysis:
\begin{eqnarray}
&&f_t=\dfrac{1}{2\pi}\sum_{k=-\infty}^\infty (f_t)_k e^{ik\theta}, 
\\
&&\mathscr{N}[f_t]=\dfrac{1}{2\pi}\sum_{k=-\infty}^\infty \mathscr{N}_k[f_t]
e^{ik\theta},\\
&&w=\dfrac{1}{2\pi}\sum_{k=-\infty}^\infty w_k e^{ik\theta}. 
\end{eqnarray}
Using Eq.~(\ref{eq:N}), we then get
\begin{eqnarray}
&&\mathscr{N}_k[f_t]=-\dfrac{kK}{2}\Big(r[f_t](-\tau)e^{-i(\alpha-\omega_0\tau)}(f_t)_{k+1}(0)-r^\ast[f_t](-\tau)e^{i(\alpha-\omega_0\tau)}(f_t)_{k-1}(0)\Big).
\nonumber \\\label{eq:N_k}
\end{eqnarray}
Note that we have
$(f_t)_0=0$, so that Eq.~(\ref{eq:decomposition}) gives $w_0=0$;
this feature is a major difference with respect to a similar kinetic equation, the Vlasov equation \cite{Crawford:1995-1}. 
By symmetry on the unstable manifold, we have~\cite{Crawford:1995-1,Guo:2013} $w_1=O(|A|^2 A)$ and for $k>1$, $w_k=O(A^k)$.
From Eq.~(\ref{eq:decomposition}), we get $(f_t)_1=A p+w_1$ and for $k\neq\pm 1$, $(f_t)_k=w_k$. 
The amplitude $A$ may be related to the order
parameter close to the bifurcation by $r= A^\ast+O(|A|^2 A^\ast)$.
Using
Eq.~(\ref{eq:decomposition}), we obtain via the  projection
$(q,(\ref{eq:ft-differential-equation-1}))_\tau$ and
$(\ref{eq:ft-differential-equation-1})-\left
((q,(\ref{eq:ft-differential-equation-1}))_\tau p+ \mathrm{c.c.}\right )$ the time evolution
of $A(t)$ and $w$ as
\begin{eqnarray}
\dot{A}&=&\lambda A+(q,\mathscr{F}[f_t])_\tau \nonumber \\
&=&\lambda A+\int {\rm d} \omega~ \widetilde{\psi}^\ast_1\mathscr{N}_1[f_t],
\label{eq:Aequation-kinetic-equation}
\\
\dot{w}&=&\mathscr{D}w+\mathscr{F}[f_t]-\left (p\int {\rm d} \omega~ \widetilde{\psi}^\ast_1\mathscr{N}_1[f_t]+\mathrm{c.c.}\right ),
\label{eq:Wequation-kinetic-equation}
\end{eqnarray}
where $\mathrm{c.c.}$ stands for complex conjugate. Here, in arriving at
the last equation, we have used $(q,\mathscr{F}[f_t])_\tau=\int {\rm
d}\omega {\rm
d}\theta~q(0)^\ast \mathscr{N}[f_t]=\int {\rm
d}\omega {\rm
d}\theta~\widetilde{\psi}^\ast_1(\omega)e^{-i\theta}\mathscr{N}[f_t]$.
From~(\ref{eq:N_k}), we get at first order 
\begin{eqnarray}
&&\mathscr{N}_1[f_t]=-A|A|^2\dfrac{K}{2}
r[p^\ast](-\tau)e^{-i(\alpha-\omega_0\tau)}w_{2,0}(0)+O(A|A|^4), 
\label{eq:N1-ft}
\end{eqnarray}
where we have denoted the leading order of the second harmonic $w_{2}=A^2
w_{2,0}+O(A^2|A|)$.

\begin{figure}[]
\begin{center}
\includegraphics[width=0.6\textwidth]{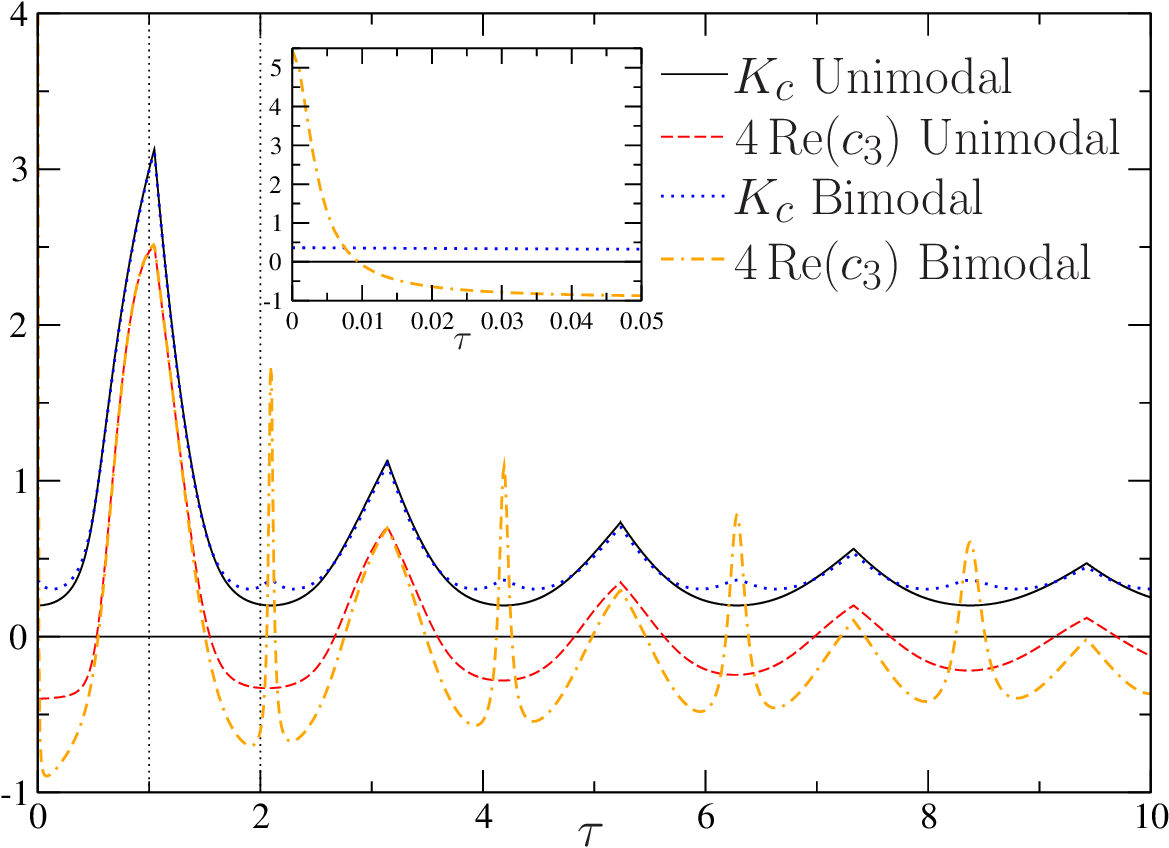}
\caption{The figure shows for $\alpha=0$ the stability region of the incoherent state for the Lorentzian
frequency distribution~(\ref{eq:lorentzian-gw}) (Unimodal) with $\Delta=0.1$,
$\omega_0=3$ and also for the case of sum of two Lorentzians (Bimodal),
Eq.~(\ref{eq:bilo}) with $\Delta=0.1,\omega_0=3$ and $\omega_c=0.09$. For $K>K_c$, the incoherent
state is unstable. The sign of $4\Re(c_3)$ 
determines the super/sub-critical nature of the bifurcation from the
incoherent state as $K \to K_c^+$: a positive (respectively, a negative)
sign implies a subcritical (respectively, a supercritical) bifurcation. The vertical dotted lines at $\tau=1$
and $\tau=2$ are where the bifurcation simulations were performed in
Fig.~4 of Ref.~\cite{Yeung:1999} for the case of the Lorentzian
$g(\omega)$, Eq.~(\ref{eq:lorentzian-gw}), with the same parameters as
us; the positive/negative sign of
$\Re(c_3)$ is fully consistent with the observed sub/super-critical
bifurcations. 
It is known that for $\tau=0$ and for the distribution~(\ref{eq:bilo}) with $\Delta/\sqrt{3}<\omega_c<\Delta$, one has
subcritical bifurcation \cite{Crawford:1994,Martens:2009}; we here see
(the inset is a zoom for small delay for the bimodal case) that on introducing even a small
amount of delay (here $\tau\gtrsim 0.01$), the bifurcation becomes
supercritical. This last prediction is explicitly verified in Fig.~\ref{fig:simu}.}
\label{fig:c3}
\end{center}
\end{figure}

Using the second harmonic of Eq.~(\ref{eq:Wequation-kinetic-equation})
and $\dot{(A^2)}=2A\dot{A}=2A^2\lambda+O(|A|^2A^2)$ gives for
$\varphi\neq 0$, $w_{2,0}=w_{2,0}(0)e^{2\varphi\tau}$. The
equation for $\varphi=0$ with Eq.~(\ref{eq:N_k}) for $k=2$ gives
\begin{eqnarray}
2 A^2\lambda w_{2,0}(0)=A^2 \Big (-2i\omega w_{2,0}(0)+ K r^\ast[p] (-\tau)e^{i(\alpha-\omega_0\tau)}p(0) \Big )
+O(|A|^2A^2).
\label{eq:UM:w2}
\end{eqnarray}
We thus get
\begin{equation}
w_{2,0}=\left (\dfrac{K}{2}\right )^2\dfrac{g(\omega)}{(\lambda+i\omega)^2}e^{2i(\alpha-\omega_0\tau)}e^{-2\lambda\tau}e^{2\varphi\tau}.
\label{eq:w2}
\end{equation}
Plugging Eqs.~(\ref{eq:w2}), (\ref{eq:N1-ft}), and
(\ref{eq:Aequation-kinetic-equation}) in
Eq.~(\ref{eq:Aequation-kinetic-equation}), we obtain the
desired normal form for the time evolution of $A(t)$:
\begin{equation}
\dot{A}=\lambda A+c_3|A|^2A+O(|A|^4A),
\label{eq:kinetic-equation-normal-form}
\end{equation}
where the cubic coefficient $c_3$ is given by
\begin{eqnarray}
&&c_3=-\dfrac{
K^3}{8}\dfrac{e^{-2\lambda\tau-\lambda^\ast\tau+i(\alpha-\omega_0\tau)}}{\Lambda'(\lambda)}\int
{\rm d}\omega~\dfrac{g(\omega)}{(\lambda+i\omega)^3} \nonumber 
\\ &&=-\dfrac{K^2}{4}
e^{-2\lambda_r\tau}\dfrac{\int{\rm
d}\omega~\dfrac{g(\omega)}{(\lambda+i\omega)^3}}{\int{\rm
d}\omega~\dfrac{g(\omega)}{(\lambda+i\omega)^2}+\tau\int{\rm d}\omega~\dfrac{g(\omega)}{(\lambda+i\omega)}}
\nonumber 
\\&&=-\dfrac{K^2}{4}
e^{-2\lambda_r\tau}\dfrac{\int{\rm
d}\omega~\dfrac{g(\omega)}{(\lambda+i\omega)^3}}{\int{\rm
d}\omega~\dfrac{g(\omega)}{(\lambda+i\omega)^2}+\tau\dfrac{2}{K}e^{\lambda\tau-i(\alpha-\omega_0\tau)}}
\\&&\overset{K\to K_c^+}{\longrightarrow}-\frac{K_c^2}{8}\left (\pi g''(\lambda_{\rm
i})-i\operatorname{PV}\int\mathrm{d}\omega~\frac{g''(\omega)}{\lambda_i+\omega}\right )
\nonumber \\
&&\times\Bigg[\operatorname{PV}\int\mathrm{d}\omega~\frac{g'(\omega)}{\lambda_{\rm{i}}+\omega}-\frac{2\tau}{K_c}\cos(\alpha-(\omega_0+\lambda_{\rm{i}})\tau)\nonumber
\\
&&\qquad-i \left (\pi
g'(\lambda_{\rm
i})-\frac{2\tau}{K_c}\sin(\alpha-(\omega_0+\lambda_{\rm{i}})\tau)\right )\Bigg]^{-1}.
\label{eq:kinetic-equation-c3}
\end{eqnarray}
Here, we have used Eq.~(\ref{eq:dispersion-kinetic-equation}), together
with the property that $g(\omega)=g(-\omega)$, to obtain
the last equality. The sign of $\Re(c_3)$ given by the last equation
gives the nature of the bifurcation as $K \to K_c^+$. Note that contrary
to similar unstable manifold analysis
\cite{Crawford:1995,Crawford:1995-1,Barre:2016} $c_3$ is not diverging
as $\lambda\to 0^++\lambda_{\rm{i}}$, which validates formally the asymptotic
analysis. Equations~(\ref{eq:ft-Kc-1}) and
(\ref{eq:kinetic-equation-c3})
suggest that at bifurcation, the effects of changing $\tau$ at a fixed
$\alpha$ are the same as those from changing $\tau$ at a fixed
$\alpha$ keeping $\alpha - \omega_0 \tau$ constant.
Interestingly, the sign of $\Re(c_3)$ predicted by our analysis, which determines the nature of
bifurcation, shows oscillations with delay (see
Fig.~\ref{fig:c3}) that agree qualitatively with what is observed in
many oscillator systems with delayed coupling or control
\cite{vanderpol:2014,delayed_control_2007,delay_feedback_2014}.

\begin{figure}[]
\begin{center}
\includegraphics[width=0.6\textwidth]{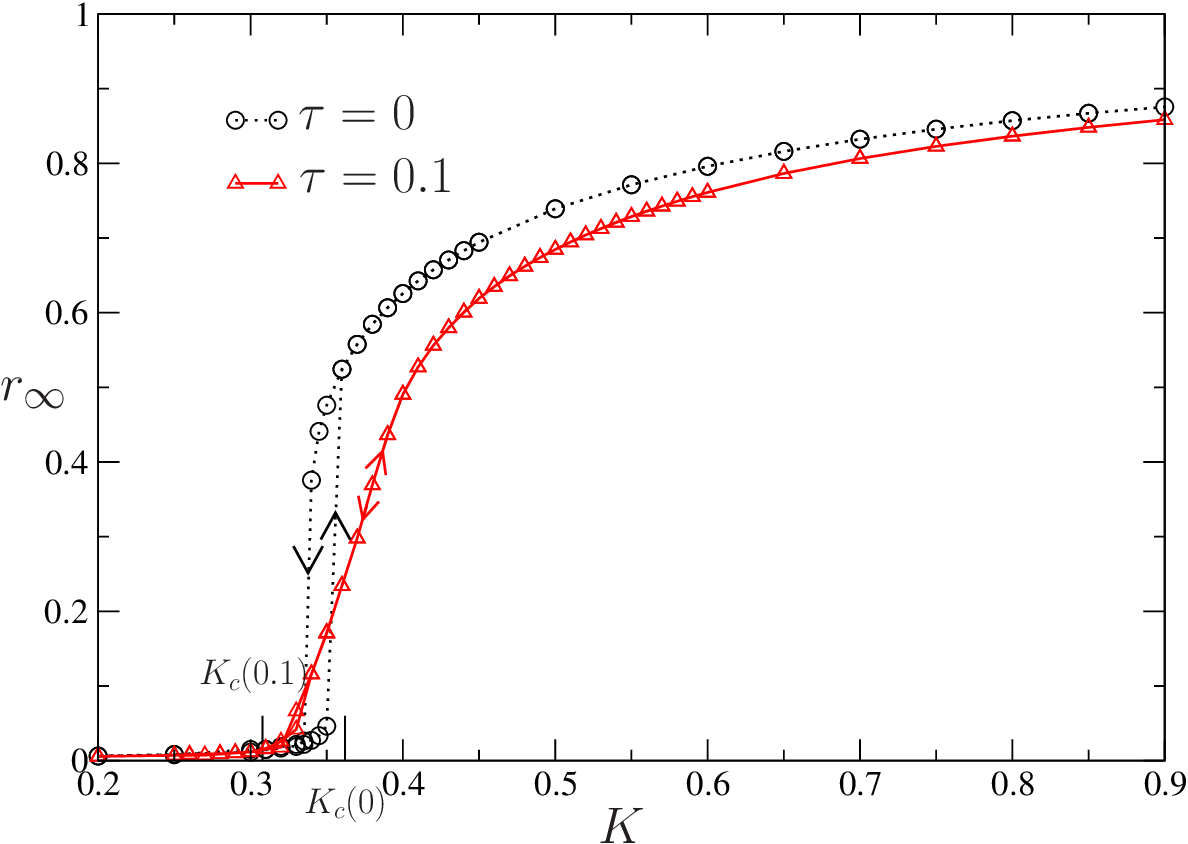}
\caption{The figure shows the variation of the order parameter $r$ as a
function of the coupling constant $K$, showing in particular the
bifurcation behavior implied by Fig.~\ref{fig:c3} for the bimodal
Lorentzian distribution, Eq.~(\ref{eq:bilo}), with
$\Delta=0.1,\omega_0=3, \omega_c=0.09$, and for two values of $\tau:$
$\tau=0$ and $\tau=0.1$. The data are obtained via numerical integration
of the dynamics~(\ref{eq:eom}) for number of oscillators $N=64384$ and
with timestep $\delta t=10^{-2}$. 
For each value $K$, we run a simulation for a total time $t=2600$ and
compute $r_\infty$ as the averaged of $|r|(t)$ for $t>1000$. We make the
end state of a simulation for a given value of $K$ as the initial state
of the run for the next value of $K$. We first increase $K$, as in $K\to
K+\delta K$, with $\delta K=0.1$ (or $0.05$/$0.5$ close/far from the bifurcation),
from low to high values, and then decreases it according to $K\to
K-\delta K$. By this procedure, we clearly differentiate the subcritical
bifurcation expected for $\tau=0$ (hysteresis behavior) from the
supercritical one expected at $\tau=0.1$ (no hysteresis). We also show the position of critical coupling $K_c(\tau)$ predicted in Fig.~\ref{fig:c3}.}
\label{fig:simu}
\end{center}
\end{figure}

For a fixed $\omega_0$ and by varying $\tau$, one may plot the sign of
$c_3$ by computing at criticality $K=K_c(\tau)$ and
$\lambda_c(\tau)=0^++i\lambda_{\rm i}(\tau)$. The result for $\alpha=0$
is shown in Fig.~\ref{fig:c3} for the case of the Lorentzian frequency
distribution, Eq.~(\ref{eq:lorentzian-gw}), and also for the case of a
sum of two Lorentzians given by
\begin{equation}
g(\omega)=\dfrac{\Delta}{\pi}\left(\dfrac{1}{(\omega-\omega_c)^2+\Delta^2}+\dfrac{1}{(\omega+\omega_c)^2+\Delta^2}\right),
\label{eq:bilo}
\end{equation}
where $\Delta$ is the width parameter of
each Lorentzian and $\pm \omega_c$ their center frequencies ($g$ has two
separated maxima for $\omega_c>\Delta/\sqrt{3}$). We show in
Fig.~\ref{fig:simu} that as predicted in Fig.~\ref{fig:c3} (inset) via
the sign of $\Re(c_3)(\tau)$, a very small delay (here $\tau=0.1$) can
suppress the subcritical bifurcation present with a bimodal distribution
in the absence of delay and turn it into a supercritical bifurcation. In effect, a subcritical bifurcation means that in the bifurcation regime, a small change in the value of the coupling $K$ leads to a large change in the value of the order parameter, that is, an incoherent state becomes with a small change of $K$ a synchronized state, and vice versa. On the contrary, a supercritical bifurcation implies that a small change in $K$ leads to only a small change in the order parameter, so that tuning of $K$ leads to a continuous change of the incoherent into a synchronized state, and vice versa. In the former case of a subcritical bifurcation, it is well known that an adiabatic tuning of $K$, in which $K$ is tuned in time at a rate slow enough  that the system is very close to the stationary state at every instant, and concomitant monitoring of $r$ leads to a hysteresis behavior of $r$ as a function of $K$~\cite{Strogatz-book}.

While Fig.~\ref{fig:c3} represents the results for Lorentzians, it is evident from the nature of the expression~(\ref{eq:kinetic-equation-c3}) that it would be quite a daunting task to make just on the basis of this expression general remarks on the nature of bifurcation for general frequency distributions, and every distribution has to be investigated on a case-by-case basis.

It may be noted that for some special values of delay $\tau=\tau_n$
satisfying $\omega_0\tau_n=2n\pi$ for $n\in \mathbb{Z}$, presence of
delay in the cosine term of Eq. (\ref{eq:eom-1}) has no effect. In this
case, if the eigenvalue triggering the instability of the incoherent
state is real (for the distribution~(\ref{eq:bilo}), it corresponds to
$|\omega_c|<\Delta$), it will be of multiplicity two, so that our derived 
two-dimensional unstable manifold is still valid. However, if there is a
pair of complex eigenvalues (for the distribution~(\ref{eq:bilo}), it
corresponds to $|\omega_c|>\Delta$), each one will have a multiplicity
of two, and consequently, one should consider instead a four-dimensional
unstable manifold, as done for $\tau_0=\tau=0$ in
Ref.~\cite{Crawford:1994-1}. For $\tau\neq \tau_n$, there is a pair of
complex eigenvalue of multiplicity one, so that our two-dimensional
unstable manifold expansion holds good.

For the Lorentzian distribution, (\ref{eq:lorentzian-gw}), one may check
that the normal form obtained from the OA-reduced-dynamics,
Eq.~(\ref{eq:OA-normal-form}), and the
kinetic equation, Eq.~(\ref{eq:kinetic-equation-normal-form}), are the same. 

For generic $g(\omega)$, we may decompose $F_t=F_{\rm{st}}+f_t$ as in Eq.
(\ref{eq:f-expansion}), with Fourier coefficients
$(F_t)_n/(2\pi)=g(\omega)/(2\pi)(\alpha_t)_n$, where the $(\alpha)_n$'s
are the Fourier coefficients on the unstable manifold. Using
Eq.~(\ref{eq:f-normalization}), we get $(\alpha_t)_0=1$. From Eqs.
(\ref{eq:psi_1}) and
(\ref{eq:decomposition}), we get
\begin{equation}
(\alpha_t)_1=A\dfrac{K}{2}
\dfrac{e^{-\lambda\tau+i(\alpha-\omega_0\tau)+\lambda\varphi}}{\lambda+i\omega}+O(A|A|^2).
\label{eq:alpha_t}
\end{equation}
Notice that expression \eqref{eq:alpha_t} explicitly satisfies the assumption made by the OA ansatz that $\alpha(\omega,t)\to 0$ as $\Im(\omega)\to -\infty$. We cannot prove within the current framework the validity of the OA assumption that $|\alpha(\omega,t)|<1$.
Similarly, Eqs. (\ref{eq:w2}) and (\ref{eq:decomposition}) give
\begin{equation}
(\alpha_t)_2=\left (A\dfrac{K}{2} \dfrac{e^{-\lambda\tau+i(\alpha-\omega_0\tau)+\lambda\varphi}}{\lambda+i\omega}\right )^2+O(A^2|A|^2).
\end{equation}
As in Eq. (\ref{eq:UM:w2}), we may write equations for the $k>1$ Fourier modes with $w_{k,0}=A^k
w_{k,0}+O(A^k|A|)$, obtaining
\begin{eqnarray}
&&w_{k,0}=w_{k,0}(0)e^{k\varphi\tau},\\
&& k  (\lambda+i\omega)w_{k,0}(0)= \dfrac{k K}{2} r^\ast[p](-\tau)e^{i(\alpha-\omega_0\tau)}w_{k-1,0}(0)+O(|A|^2 ),
\label{eq:UM:Nk}
\end{eqnarray}
where we have used $\dot{(A)^k}=k\lambda A^k+O(A^k|A|^2)$, and the fact
that the dominant contribution in Eq.~(\ref{eq:N_k}) always involves for $k>1$ the $f_{k-1}$ term and not $f_{k+1}$.
By induction, we deduce for $k\geq 0$ that
\begin{equation}
(\alpha_t)_k=(\alpha_t)_1^k+O(A^k|A|^2),
\label{eq:UM:oa}
\end{equation}
and by taking the complex conjugate of the last equation, we obtain the corresponding equation for $k<0$. 
These equations clearly show that the unstable manifold has exactly the same form as the OA manifold close to the bifurcation. Note that this feature is
valid both in the absence and presence of delay. It is worthwhile to
mention that the OA ansatz fails on adding a second harmonic (that is,
with the form of interaction $\sim K\sin\theta+ J\sin(2\theta)$) to
Eq.~(\ref{eq:eom-kuramoto}), and even the relation (\ref{eq:UM:oa})
obtained with Eq.~(\ref{eq:UM:Nk}) is also not valid in this case.
In fact in this case (without delay) the unstable manifold has been shown to be singular \cite{Crawford:1995,Crawford:1999}. Nonetheless, the unstable manifold reduction still provided precious informations on the bifurcation. The singularities denote a profound change in the nature of the problem.
Studying how Eq.~(\ref{eq:UM:Nk}) is modified by the addition of a
second harmonic or noise could be a starting point for investigations into
generalizations of the OA ansatz. 
%%%%%%%%%%%%%%%%%%%%%%%%%%%%%%%%%%%%%%%%%%%%%%%%%%%%%%%%%%%%%%%%%%%%%%%%%%%%%%%%%%%%%%
\section{Conclusions and perspectives}
\label{sec:conclusions}

In this work, we analyzed in detail the consequences of a time delay in
the interaction in the case of the Kuramoto model of globally-coupled oscillators with
distributed natural frequencies, for generic choice of the frequency
distribution $g(\omega)$. We derived as a function of the
delay exact results for the stability boundary between the
incoherent and the synchronized state and the nature in which the latter
bifurcates from the former at the critical
point. Our results are obtained in two independent ways: one, by
considering the kinetic equation for the time evolution of the
single-oscillator distribution, and two, by considering for the
specific choice of a Lorentzian distribution a reduced
equation for the order parameter derived from the kinetic equation by invoking the celebrated Ott-Antonsen ansatz. In either case, the
incoherent state, in which the oscillators are completely
unsynchronized, is a stationary solution for all values of the coupling
constant $K$ between the oscillators, but which is linearly stable only
below a critical value $K_c$ of the coupling. To
examine how a stable synchronized state bifurcates from the incoherent
state as the coupling crosses the value $K_c$, we employed an unstable manifold
expansion of perturbations about the incoherent state in the vicinity of the bifurcation, which we
applied both to the kinetic equation and to the corresponding Ott-Antonsen-reduced dynamics. We found that the nature of the bifurcation is determined by the sign of the coefficient of the
cubic term in the equation describing the amplitude dynamics of the
unstable modes in the regime of weak linear
instability, namely, as $K \to K_c^+$. Remarkably, we found that the amplitude
equation derived from the kinetic equation has the same form as that obtained
from the OA-ansatz-reduced dynamics for the particular case of a Lorentzian
$g(\omega)$, thus confirming the power and general applicability of the
OA ansatz. Moreover, quite interestingly, we found that close to the bifurcation, the unstable manifold
has the same form as that of the OA manifold. This may have important bearings on their inter-relationship to be unravelled in future.
As an explicit physical effect of the presence of delay, we demonstrated with our exact results
that for a sum of two Lorentzians as a representative
example of a bimodal frequency distribution, while absence of
delay leads to a bifurcation of the synchronized from the incoherent
state that is subcritical, even a small amount of delay changes completely the nature of the bifurcation and makes it supercritical. 

%It remains as an exercise for future work to
%study how our analysis and results are modified on considering
%additional harmonics in the interaction between the
%Kuramoto oscillators, for which the present work may serve as a genesis.

\begin{acknowledgements}
This work was initiated while DM was affiliated to Laboratory J.A. Dieudonn\'e, 
Universit\'{e} C\^{o}te d'Azur, Nice, France and was finalized with DM
being affiliated to Los Alamos National Laboratory (LANL). DM gratefully
acknowledges the support of the U.S. Department of Energy through the
LANL/LDRD Program and the Center for Non Linear Studies, LANL.
The paper was written up during the visit of DM and SG to the
International Centre for Theoretical Physics -- South American Institute
for Fundamental Research, S\~{a}o Paulo, Brazil in May 2018 and during
SG's extended stay at
the Universidade Federal de S\~{a}o Carlos and the Centro de Pesquisa em
\'{O}ptica e Fot\^{o}nica, S\`{a}o Carlos, Brazil during June 2018. The
authors thank these institutions for warm hospitality and financial support.
\end{acknowledgements}

\end{document}